\documentclass{SCGE}
\usepackage[colorlinks,linkcolor=blue,citecolor=blue,urlcolor=blue]{hyperref}

\usepackage[latin1]{inputenc}
\usepackage{rotating}
\usepackage{footnote}
\newcolumntype{C}[1]{>{\centering\let\newline\\\arraybackslash\hspace{0pt}}m{#1}}

\usepackage{times}
\usepackage{newtxtext,newtxmath}
\usepackage{graphicx}	
\usepackage{amsmath}	
\usepackage{amssymb}	
\usepackage{xspace}
\usepackage{bigints}
\usepackage{siunitx}
\usepackage{amsthm}
\usepackage{mathtools}
\usepackage[normalem]{ulem}
\usepackage{float}
\usepackage{color}

\newcommand{\NHn}{NH$_3$}

\newcommand{\Hn}{H$_2$}

\newcommand{\HCnN}{HC$_{3}$N}

\newcommand{\Msun}{$M_\odot$}

\newcommand{\kms}{{km~s$^{-1}$}}

\usepackage{CJK}

\let\citedash\relax
\makeatletter \providecommand{\citedash}{\hbox{-}\penalty\@m}
\makeatother

\begin{document}

\begin{picture}(0,0){\rm
\put(0,-20){\makebox[160truemm][l]{\bf {\sanhao\raisebox{2pt}{.}}
Article  {\sanhao\raisebox{1.5pt}{.}}}}}
\put(0,-34){\jiuwuhao {\textcolor[rgb]{0.5,0.5,0.5}{\sf 
}}}
\end{picture}

\def\bm{\boldsymbol}

\def\dl{\displaystyle}
\def\du{\end{document}}
\def\d{{\rm d}}
\def\e{{\rm e}}
\def\i{{\rm i}}

\Year{2020} %
\Month{October} %
\Vol{59} %
\No{1} %
\BeginPage{1} %
\AuthorMark{{\rm J. XIE}, et al.}  
\DOI{} 
\ArtNo{000000}

\title[TMRT IRDCs]{The TMRT K Band Observations towards 26 Infrared Dark Clouds: {\NHn}, CCS, and {\HCnN}}

\author[1,2,3]{Jinjin Xie}{}
\author[3,4,5]{Gary~A.~Fuller}{}
\author[1,2,6]{Di Li*}{}
\footnote{*Corresponding author: dili@nao.cas.cn}
\author[1]{Longfei Chen}{}
\author[1,2]{Zhiyuan Ren}{}
\author[1,2]{\\Jingwen Wu}{}
\author[1,2]{Yan Duan}{}
\author[7]{Junzhi Wang}{}
\author[7]{Juan Li}{}
\author[8]{Nicolas Peretto}{}
\author[7]{Tie Liu}{}
\author[7]{Zhiqiang Shen}{}

\address[{\rm1}]{National Astronomical Observatories, Chinese Academy of Sciences, Beijing 100101, China;}
\address[{\rm2}]{University of Chinese Academy of Sciences, Beijing 100049, China;}
\address[{\rm3}]{Jodrell Bank Centre for Astrophysics, Department of Physics and Astronomy, University of Manchester, Manchester M13 9PL, the United Kingdom;}
\address[{\rm4}]{Intituto de Astrof\'isica de Andalucia (CSIC), Granada 18008, Spain;}
\address[{\rm5}]{I. Physikalisches Institut, University of Cologne, K\"oln 50937, Germany;}
\address[{\rm6}]{NAOC-UKZN Computational Astrophysics Centre, University of KwaZulu-Natal, Durban 4000, South Africa;}
\address[{\rm7}]{Shanghai Astronomical Observatory, Shanghai 200030, China}
\address[{\rm8}]{School of Physics \& Astronomy, Cardiff University, Cardiff CF24 3AA, the United Kingdom;}

\maketitle \vspace{-3.5mm}{\footnotesize\begin{center} Received January 1, 2016; accepted January 1, 2016; published online January 1, 2016
\end{center}}\vspace*{-5mm}

\begin{center}
\rule{16.5cm}{0.4pt}
\parbox{16.5cm}
{\begin{abstract}
We present one of the first Shanghai Tian Ma Radio Telescope (TMRT) K Band observations towards a sample of 26 infrared dark clouds (IRDCs). We observed the (1,1), (2,2), (3,3), and (4,4) transitions of {\NHn} together with CCS (2$_{1}$-->1$_{0}$) and {\HCnN} $J\,=$2-1, simultaneously. The survey dramatically increases the existing CCS-detected IRDC sample from 8 to 23, enabling a better statistical study of the ratios of carbon-chain molecules (CCM) to N-bearing molecules in IRDCs. With the newly developed hyperfine group ratio (HFGR) method of fitting NH$_3$ inversion lines, we found the gas temperature to be between 10 and 18\,K. The column density ratios of CCS to {\NHn} for most of the IRDCs are less than 10$^{-2}$, distinguishing IRDCs from low-mass star-forming regions. We carried out chemical evolution simulations based on a three-phase chemical model {\sc NAUTILUS}. Our measurements of the column density ratios between CCM and {\NHn} are consistent with chemical evolutionary ages of $\lesssim$10$^{5}$\,yr in the models. Comparisons of the data and chemical models suggest that CCS, {\HCnN}, and {\NHn} are sensitive to the chemical evolutionary stages of the sources.

\end{abstract}}
\end{center}\vspace*{-0.6cm}

\begin{center}
\parbox{16.5cm}
{\bf\jiuhao Star formation, infrared dark clouds, chemical evolution}
\end{center}

\begin{center}
{}
\end{center}

\textwidth=178truemm \textheight=236truemm

\wuhao\vspace*{1.5mm}


\renewcommand{\baselinestretch}{1.08} \baselineskip 12.2pt\parindent=10.8pt

\section{Introduction}\label{sec:intro}


The timescale of chemical evolution in massive star forming regions and its relation to the onset of massive star formation are still unclear despite its importance. The earliest stage of high-mass star formation is considered to take place in the regions with low temperatures (<25\,K), high column densities ($\sim$10$^{23}$-10$^{25}$\,{cm$^{-2}$}), and high volume densities ($\gtrsim$10$^{5}$\,{cm$^{-3}$})\cite{rathborne2006}. Infrared Dark Clouds (IRDCs), seen as silhouette against infrared emission, are thus considered to be representative of the initial conditions for massive star formation and the formation of the associated stellar clusters \cite{peretto2009,motte2018}.

Molecular lines are powerful diagnostic tools for studying the physical conditions and chemical evolutionary stages of star-forming regions \cite{caselli2012,vanDishoeck2018}. The combination of the carbon-chain molecule (CCM) and {\NHn}, in particular, has been widely adopted to explore low-mass star-forming regions\cite{hirahara1992, hirota2009, suzuki2014}. In the early stage of star formation, CCMs tend to become deficient \cite{sakai2008,hirota2009}, while the abundances of N-bearing molecules such as {\NHn} tend to increase \cite{suzuki1992,tatemastu2017}. {\NHn} is known as an excellent dense gas tracer as well as a reliable thermometer for interstellar clouds as it is sensitive to collisional excitation and insensitive to radiative excitation \cite{ho1983,walmsley1983,li2003}.
The distribution of {\NHn} is found to be centrally condensed and surrounded by CCMs in low-mass star-forming regions (e.g.,\cite{kuiper1996,ohashi1999}). The variations of the column density ratio of N-bearing species and CCMs are thus known as indicators of chemical evolution in low-mass star-forming regions \cite{suzuki1992,benson1998,suzuki2014}.

Massive star-forming regions are less explored in studies of astrochemistry \cite{vasyunina2014}. The first attempt to apply such "chemical age indicator" to IRDCs was made with a single-pointing survey towards 16 Infrared Dark Clouds (IRDCs) using the Nobeyama Radio Observatory (NRO) 45\,m telescope. The non-detection of CCS led to the conclusion that these massive star-forming regions are more evolved than their low-mass counterparts \cite{sakai2008}. Using a 100\,m single-dish telescope (Robert C. Byrd Green Bank Telescope, GBT) and an interferometric array (the Very Large Array, VLA), CCS was mapped in a small sample of IRDCs and was found to have similar distributions surrounding {\NHn} as those in the low-mass star-forming regions \cite{devine2011, dirienzo2015}. Previous observations and modelling suggest that the low CCS abundance may be a feature intrinsic to IRDCs\cite{devine2011,vasyunina2012,dirienzo2015}.

The Shanghai Tian Ma Radio Telescope (TMRT) is a 65\,m diameter fully steerable radio telescope located in the western suburb of Shanghai, China \cite{lij2016}. The K Band receiver in TMRT can observe {\NHn}, CCS, and {\HCnN} simultaneously, reducing the uncertainties due to the observing system when determining the column density ratios of these species. In this study, we used the TMRT to conduct single-pointing observations towards 26 IRDCs from a carefully-selected sample of IRDCs (Peretto et al. in prep.) with representative parameters  and rich complementary data such as N$_{2}$H$^{+}$. From these observations, we obtained the gas temperatures, velocities, velocity dispersions, and column densities, and derived relative abundances of respective species. We also carried out chemical modelling to simulate the evolutionary tracks of  the observed relative abundances. Our observation and modeling  shed lights into the chemical evolution of these regions, at the onset of massive star formation.

\section{Sources and Observation} \label{sec:observation}

This sample of 26 IRDCs observed here has been selected from the catalogue of IRDCs identified by \cite{peretto2009}. The sources in the sample were selected to have kinematic distances between 3 and 5\,kpc \cite{jackson2006} and a range of morphologies from round to filamentary as seen in the 8\,$\mu$m absorption and {\it Herschel} column density images \cite{peretto2016}. The sources span a range of masses from 258 to $2.0\times10^4$\,{\Msun} and sizes from 1 to 11\,pc. A more extensive description of the sample is given in Peretto et al. (in prep).

We performed single-pointing observations towards 26 IRDCs with the TMRT K band receiver (18--26.5\,GHz) in January and March 2019. Every two hours, we observed a calibration source selected amongst 3C273, DR21, 3C279, 3C453, and NRAO530, depending on which was near the target source, with cross scans to check the pointing of the telescope. The pointing accuracy was better than 10\% beam size. The observations were conducted in the position-switching mode. The off-source positions were chosen to be +10 arcmin in azimuth away from the source, and were free from emission.

A Field-Programmable Gate Array (FPGA)-based spectrometer following the design of the Versatile GBT Astronomical Spectrometer (VEGAS) was employed as the Digital backend system (DIBAS) \cite{bussa2012,wu2017}. We adopted mode 22 in DIBAS with 8 spectral windows to cover the following molecular lines: {\NHn}(1,1) [23.6944955\,GHz \cite{lovas2003}], {\NHn}(2,2) [23.7226333\,GHz \cite{lovas2003}], {\NHn}(3,3) [23.8701292\,GHz \cite{poynter1975}], {\NHn}(4,4) [24.1394163\,GHz \cite{poynter1975}],
CCS(2$_{1}$-->1$_{0}$) [22.344033\,GHz\cite{yamamoto1990}], and {\HCnN}(2-->1) [18.196279\,GHz\cite{thorwirth2000}] simultaneously. Each of the eight subbands has a bandwidth of 23.4\,MHz and 16384 channels. The spectral resolution is 1.431\,kHz, corresponding to a velocity resolution of 0.024\,{\kms} at 18\,GHz and 0.018\,{\kms} at 24\,GHz. The half power beam widths (HPBWs) of the TMRT beam at our observed frequencies are about 48" at K Band \cite{hu2021}, while for {\HCnN} the HPBW is around 54"\cite{lij2016}. The main beam efficiency is 45\% at the K band \cite{wang2017}. The intensities were calibrated by injecting periodic noise and the accuracy of the intensity calibration is 20\%. The system temperatures are around 190\,K.

The observations were made at the peak column density positions derived from the {\it Herschel} images of the regions\cite{peretto2016}. The physical parameters of the observed IRDCs are listed in Table \ref{tab:sources_parameters}, including the source names, positions, distances, and the average {\Hn} column densities. Each molecular transition observation had two polarisations, which, after visual inspection and the removal of any bad scans, were averaged to reduce the rms noise levels. The on and off source integration time were 120 seconds and were repeated five times for each source, reaching a typical rms noise of 0.07\,K in the 143\,kHz ($\sim0.2$\,\kms) wide channel, which is used to analyse the CCS and {\HCnN} observations, and 0.23\,K in the 1.43\,kHz channels for \NHn. The rms noises of these observations vary from 0.04\,K to 0.19\,K for CCS and {\HCnN} observations and 0.09\,K to 0.62\,K for {\NHn} observations, to which we have included the uncertainties which are related to the rms noise in the analysis for each source.

The software package {\sc CLASS/GILDAS}\footnote{https://www.iram.fr/IRAMFR/GILDAS/} \cite{guilloteau2000} was used to reduce the observed molecular line data and to plot the spectra. Linear baseline subtractions were used for all the spectra.

\clearpage
\begin{tablehere}
\vspace{5mm}\footnotesize
\caption{Physical Parameters of the Observed IRDCs.}\label{tab:sources_parameters}
\begin{center} \doublerulesep 0.2pt \tabcolsep 4.2 pt
\begin{tabular}{lcc c c}
\hline
 & R.A.(J2000) & Decl.(J2000)  & Distance$^{(a)}$ & <$N_{H_{2}}$>$^{(a)}$ \\
Source Name & (hh:mm:ss) & ($^\circ$ ' ") & (kpc) & (10$^{22}$\,cm$^{-2}$)\\
\hline
SDC18.624-0.070 & 18:25:10.0 & -12:43:45 &   3.5 & 5.7 \\
SDC18.787-0.286 & 18:26:19.0 & -12:41:16  & 4.3  & 6.4 \\
SDC18.888-0.476 & 18:27:09.7 & -12:41:32 &  4.3  & 9.0\\
SDC21.321-0.139 & 18:30:32.1 & -10:22:50 &  4.2  & 4.6 \\
SDC22.724-0.269 & 18:33:38.3 & -09:11:55 &  4.4 & 6.0\\    
SDC23.066+0.049 & 18:33:08.3 & -08:44:53  & 5.1 & 5.4\\
SDC23.367-0.288 & 18:34:53.8 & -08:38:00  &   4.6 & 11.3\\
SDC24.118-0.175 & 18:35:52.6 & -07:55:06 &  4.7  & 5.4\\
SDC24.433-0.231 & 18:36:41.0 & -07:39:20 &  3.8 & 7.0\\
SDC24.489-0.689 & 18:38:25.7 & -07:49:36  &  3.3 & 3.9\\
SDC24.618-0.323 & 18:37:22.4 & -07:32:18 &  3.0 & 4.5\\
SDC24.630+0.151 & 18:35:38.2 & -07:18:35 &  3.5  & 6.0\\
SDC25.166-0.306 & 18:38:13.0 & -07:03:00  & 3.9 & 5.9\\
SDC25.243-0..447 & 18:38:57.1 & -07:02:20 &  3.8  & 4.4\\
SDC26.507+0.716 & 18:37:07.9 & -05:23:58 & 3.2  & 3.7\\
SDC28.275-0.163 & 18:43:30.3 & -04:12:45  &  4.6 & 6.3\\
SDC28.333+0.063 & 18:42:54.1 & -04:02:30  & 4.6 & 10.6\\
SDC31.039+0.241 & 18:47:03.3 & -01:33:50& 4.5  & 7.0\\
SDC34.370+0.203 & 18:53:18.9 & +01:24:54  &  3.6  & 8.1\\
SDC35.429+0.138 & 18:55:30.4 & +02:17:10 &  4.7 & 6.8\\
SDC35.527-0.269 & 18:57:08.6 & +02:09:08  & 2.9  & 5.1\\
SDC35.745+0.147 & 18:56:02.6 & +02:34:14 &  5.1 & 4.7\\
SDC38.850-0.427 & 19:03:46.8 & +05:04:03  & 2.8 & 4.9\\
SDC40.283-0.216 & 19:05:41.2 & +06:26:09  & 4.9 & 5.1\\
SDC47.061+0.257 & 19:16:41.8 & +12:39:39 & 4.6 & 4.5\\
SDC52.723+0.045 & 19:28:34.4 & +17:34:17  & 4.5  & 3.0\\
\hline
\multicolumn{5}{p{.45\textwidth}}{Notes. $^{(a)}$ The values of these physical properties are from Peretto et al. (in prep).}
\end{tabular}
\end{center}
\end{tablehere}

\section{Results}
All the 26 IRDCs were detected in {\NHn} (1,1) and (2,2) lines. {\NHn} (3,3) was detected in 17 sources, while {\NHn}(4,4) was detected in four sources. Fifteen sources were detected in CCS and nineteen in {\HCnN}, corresponding to a detection rate of 58\% and 73\%, respectively. The peak vlsr (velocity with respect to local standard of rest) and velocity width of the {\NHn} (1,1) emission were measured by fitting the full hyperfine structures of the transitions using {\sc CLASS}. Two sources, namely SDC26.507+0.716 and SDC47.061+0.257, have two velocity components much overlapped in {\NHn} profiles. Their signal-to-noise ratios of CCS and {\HCnN} are too low to distinguish with high confidence. We thus only fitted CCS and {\HCnN} with one velocity component and excluded them from further analysis.

For the CCS, the line properties were determined from a Gaussian fit to the line profile. The {\HCnN} $J=$2-1 line is split into 6 hyperfine components owing to the interaction of the nuclear electric-quadrupole moment of the nitrogen with the molecular field gradient \cite{dezafra1971}. To determine the properties of the {\HCnN} lines, the full hyperfine structure was fitted to the spectra using {\sc CLASS}. In principle, this allows the determination of the optical depth of the transition, but all the spectra were consistent with optically thin emission. The final line parameters were then determined using the fit to the hyperfine components assuming that the emission was optically thin. For the full set of spectra please see supplementary material.

\clearpage

\begin{tablehere}
\caption{Observed Parameters of {\NHn}(1,1), CCS, and {\HCnN}}\label{tab:observed_parameters}
\vspace{-1mm}\footnotesize
\begin{center} \doublerulesep 0.3pt \tabcolsep 3pt
\begin{tabular} {c c c c  c  c c  c c c }
\hline
 &  & {\NHn}(1,1) & &&  CCS  & & & {\HCnN}& \\
Source Name & T$_{mb}$ & Velocity & Width & T$_{mb}$ & Velocity & Width  & T$_{mb}$ & Velocity & Width \\
 & (K)  & (km s$^{-1}$) & (km s$^{-1}$) & (K) & (km s$^{-1}$) & (km s$^{-1}$) & (K) & (km s$^{-1}$) & (km s$^{-1}$)\\
 \hline
SDC18.624-0.070 & 1.96 & 46.14 (0.02) & 1.77 (0.05) & <0.36 & - & - &  <0.93 & - &  -  \\
 SDC18.787-0.286 & 2.21 & 65.49 (0.01) & 1.60 (0.02) & 0.12 & 64.34 (0.28) & 3.16 (1.01) & 0.17 & 65.35 (0.20) & 2.62 (0.53)\\
  SDC18.888-0.476 & 1.63 & 66.67 (0.01) & 1.74 (0.02) & <0.24  & - & - & 0.12 & 66.42 (0.25) & 4.56 (0.64)\\
 SDC21.321-0.139 & 0.97 & 66.42 (0.02) & 1.30 (0.05) & <0.18 & - & - & <0.18 & -& - \\
 SDC22.724-0.269[1] & 1.02 & 73.17 (0.01) & 1.20 (0.02) & <0.11 & - & - & <0.12 & - & -\\
  SDC22.724-0.269[2] & 0.64 & 104.43 (0.06) & 2.47 (0.08) & 0.04 & 105.06 (0.63) & 4.38 (1.35) & <0.12 & - & -\\
 SDC23.066+0.049 & 1.02 & 91.33 (0.02) & 1.91 (0.05) & 0.14 & 90.52 (0.12) & 0.76 (0.26) & 0.12 & 91.44 (0.20) & 1.03 (0.43)\\
 SDC23.367-0.288 & 1.40 & 78.46 (0.07) & 2.38 (0.06) & 0.21 & 77.89 (0.11) & 1.31 (0.24) & 0.19 & 78.29 (0.19) & 2.88 (0.46)\\
 SDC24.118-0.175 & 2.92 & 80.80 (0.03) & 1.47 (0.02) & <0.33 & - & - & 0.26 & 80.92 (0.18) & 2.66 (0.60)\\
 SDC24.433-0.231 & 3.12 & 58.68 (0.06) & 2.67 (0.02) & <0.27 & - & - & 0.35 & 58.21 (0.20) & 3.00 (0.47)\\
 SDC24.489-0.689 & 2.38 & 48.52 (0.05) & 2.17 (0.02) & 0.24 & 48.19 (0.12) & 1.96 (0.26) & 0.38 & 48.17 (0.09) & 2.90 (0.24)\\
 SDC24.618-0.323 & 1.75 & 43.20(0.03) & 1.49 (0.04) & <0.24 & - & - & 0.23 & 42.77 (0.16) & 1.39 (0.28)\\
 SDC24.630+0.151 & 1.90 & 53.14 (0.03) & 1.69 (0.03) & 0.15 & 52.33 (0.32) & 2.92 (0.77) & 0.42 & 53.57 (0.09) & 2.00 (0.16)\\
 SDC25.166-0.306 & 1.44 & 63.68 (0.03) & 1.40 (0.01) & 0.11 & 62.75 (0.29) & 2.87 (0.77) & <0.24 & -& -\\
 SDC25.243-0.447 & 1.82 & 59.10 (0.03) & 1.14 (0.02) & 0.18 & 57.73 (0.07) & 0.52 (0.13) & 0.07 & 58.69 (0.66) & 3.01 (1.76)\\
 SDC26.507+0.716[1] & 2.15 & 47.74 (0.03) & 0.56 (0.06) & 0.20 & 47.24 (0.65) & 5.14 (2.44) & 0.28 & 48.47 (0.20) & 2.10 (0.45)\\
  SDC26.507+0.716[2]$^{(a)}$ & 2.15 & 49.15 (0.09) & 1.83 (0.19) & - & - & - & - & - & -\\
 SDC28.275-0.163 & 1.26 & 80.60 (0.07) & 1.70 (0.08) & <0.45 & - & - & <0.45 & - & -\\
 SDC28.333+0.063 &2.62 & 79.71 (0.04) & 1.83 (0.02) & 0.22 & 78.82 (0.13) & 2.31 (0.29) & 0.27 & 78.20 (0.15) & 3.57 (0.41)\\
SDC31.039+0.241[1] & 2.47 & 78.22 (0.03) & 1.74 (0.04) & <0.21 & - & - & 0.12 &79.73 (0.15) & 3.55 (0.40) \\
 SDC31.039+0.241[2] & 2.32 & 97.49 (0.04)  & 1.03 (0.02)  & <0.21 & - & - & 0.12 & 98.13 (0.35) & 1.69 (0.61)  \\
 SDC34.370+0.203 & 2.57 & 58.13 (0.07) & 3.31 (0.07) & <0.57 & - &-  & <0.45 &  - & - \\
 SDC35.429+0.138 & 2.10 & 75.83 (0.03) & 1.22 (0.02) & 0.11 & 75.09 (0.16) & 2.29 (0.37) & 0.24 & 76.33 (0.06) & 1.33 (0.12)\\
 SDC35.527-0.269 & 2.54 & 45.50 (0.03) & 1.00 (0.01) & 0.19 & 44.78 (0.13) & 1.71 (0.28) & 0.32 & 45.52 (0.04) & 0.81 (0.09)\\
 SDC35.745+0.147 & 3.61 & 83.19 (0.03) & 1.79 (0.01) & 0.14 & 81.97 (0.20) & 2.06 (0.60) & 0.23 & 83.38 (0.17) & 2.90 (0.41)\\
 SDC38.850-0.427 & 2.14 & 42.33 (0.03) & 1.13 (0.02) & 0.08 & 41.33 (0.68) & 5.72 (1.39) & 0.17 & 42.59 (0.19) & 1.89 (0.35)\\
 SDC40.283-0.216 & 1.77 & 73.26 (0.05) & 3.21 (0.05) & <0.24 & - & - & 0.20 & 73.24 (0.21) & 1.97 (0.47) \\
 SDC47.061+0.257[1]$^{(a)}$ &1.01 & 55.17 (0.12) & 0.99 (0.15) & - & - & - & - & - & - \\
  SDC47.061+0.257[2] &1.01 & 56.77 (0.14) & 1.43 (0.29) & 0.08 & 56.26 (0.56) & 3.79 (0.96) & <0.18 & - & - \\
 SDC52.723+0.045 & 0.46 &44.47 (0.06) & 1.27 (0.10) & <0.18 & - & - & 0.11 & 44.03 (0.42) & 1.78 (0.66)\\
 \hline
\multicolumn{10}{p{.9\textwidth}}{Notes. The limits are determined from 3$\sigma$ rms noise level. The two velocity components detected towards some sources are indicated by [1] and [2] after the source name. $^{(a)}$ The fits refer to single velocity fit. The signal-to-noise ratios are too low for CCS and {\HCnN} in these two sources (SDC26.507+0.716 and SDC47.061+0.257) to refer to the two velocity components.} 
\end{tabular}
\end{center}
\end{tablehere}
\clearpage

\subsection{Gas and Dust Temperatures}\label{subsec:temperatures}

We use the newly developed hyperfine group ratio (HFGR)\cite{wang2020,li2013} method, which utilizes the integrated intensities of hyperfine groups to derive $T_{rot}$ and $T_{k}$. Relying solely on observed line-intensity-ratios and completely circumventing hyperfine-fitting, HFGR is impervious to the coupling between linewidth and opacity, thus more robust and accurate for a wider range of parameters than all traditional methods. A key product of the HFGR is the statistical uncertainties of the fitted temperatures based on the signal-to-noise ratio of the observed spectra \cite{wang2020}. The temperatures of the regions range between 10\,K and 18\,K.

\begin{figure}[H]
\centering
\includegraphics[width=0.50\textwidth,height=0.40\textwidth]{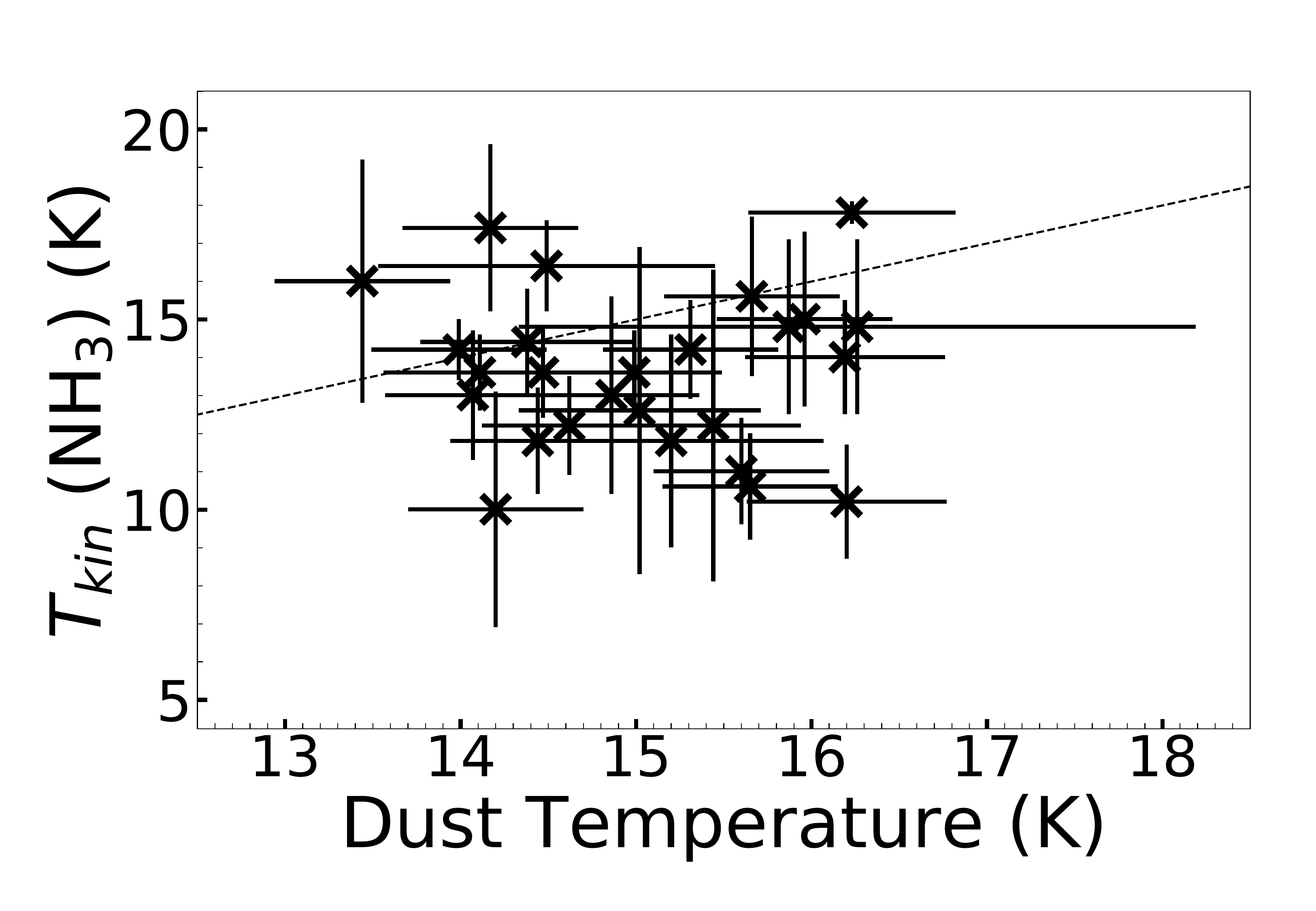}
\caption{The comparison of the dust temperatures derived from the Hi-GAL observations and the {\NHn} kinetic temperatures from this survey. The dotted slope line represents where the gas temperature equals the dust temperature. The {\NHn} kinetic temperatures are typically slightly lower than those of dust, by $\lesssim$1\,K on average.
} 
\label{fig:temperature_difference}
\end{figure}

Figure~\ref{fig:temperature_difference} compares the dust temperatures from the {\it Herschel} Far-Infrared observations\cite{peretto2016} with gas temperatures from this K Band {\NHn} survey. The dust temperatures are derived from the 160 $\mu$m and 250 $\mu$m data obtained with the {\it Herschel} Galactic plane survey Hi-GAL\cite{peretto2016}. To compare with the TMRT data, the Hi-GAL temperatures and the corresponding uncertainties are smoothed to the TMRT beam size at {\NHn} frequency. The dust temperatures are similar to the {\NHn} kinetic temperatures, but are typically $\lesssim$1\,K higher than the {\NHn} temperatures, which has been noted for IRDCs previously\cite{sokolov2017}.

\subsection{Line Widths and Velocities}\label{subsec:width_velocity}

\begin{figure}[H]
\centering
\includegraphics[width=0.50\textwidth,height=0.30\textwidth]{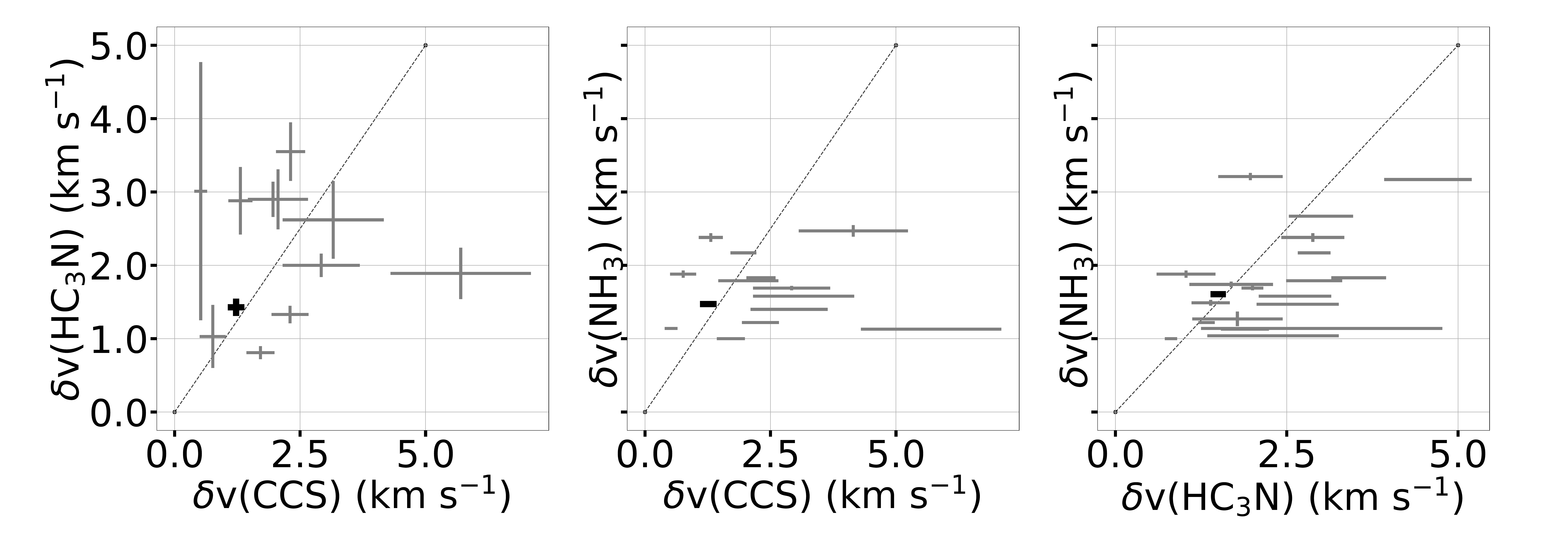}
\caption{The comparison of the line widths among CCS, {\HCnN}, and {\NHn}. From left to right panels are the comparison of line widths between {\HCnN} and CCS, {\NHn} and CCS, and {\NHn} and CCS. The gray points and errorbars represent the observed line widths and the corresponding errors. The black points and errorbars represent the weighted mean line widths and errors. The diagonal shows the equality of line widths. } 
\label{fig:weighted_linewidth}
\end{figure}

Figure~\ref{fig:weighted_linewidth} shows the comparison of line widths (FWHM, dV) of  {\NHn}, {\HCnN}, and CCS. The average line widths of all molecules are similar, suggesting the molecules are tracing similar regions.

\subsection{Column density determination}\label{subsec:column_density}

The total column density of {\NHn} is determined using {\it N}(1,1) $\times$Z/Z(1,1), where:
\begin{equation}
\begin{aligned}
Z = \sum_{i}\left(2J+1\right){\it S}\left(J\right)\exp\frac{-h\left[BJ\left(J+1\right)+\left(C-B\right)J^{2}\right]}{kT_{rot}},
\end{aligned}
\end{equation}
and 
\begin{equation}
\begin{aligned}
Z(1,1) = 3S\left(1\right)\exp\frac{-h\left[2B+\left(C-B\right)\right]}{kT_{rot}},
\end{aligned}
\end{equation}
where {\it S(J)} is the extra statistical weight of the ortho- over para-{\NHn} states and equals 2 for {\it J} = 3, 6, 9, ... and equals 1 for other {\it J}. The values for the rotational constants B and C are 298117\,MHz and 186726\,MHz, respectively\cite{pickett1998}. h and k are the Planck constant and Boltzmann constant, respectively. The column density of the {\NHn} (1,1) transition is then calculated using equation from\cite{friesen2009}:
\begin{equation}\label{eq:n11}
\begin{aligned}
{\it N}(1,1) = \frac{8\pi\nu_{0}^{2}}{c^{2}}\frac{g_{1}}{g_{2}}\frac{1}{A\left(1,1\right)}
\times \frac{1+\exp\left(-h\nu_{0}/kT_{ex}\right)}{1-\exp\left(-h\nu_{0}/kT_{ex}\right)} \int\tau\left(\nu\right)d\nu,
\end{aligned}
\end{equation}
where g is the degeneracy of the corresponding energy level. $\tau_{\nu}$ is the optical depth as a function of frequency $\nu$ determined from the hyperfine fit. The Einstein A coefficient {\it A}(1,1) is 1.68$\times$10$^{-7}$\,s$^{-1}$\cite{pickett1998}.

For CCS and {\HCnN}, assuming the lines are optically thin and are in local thermodynamic equilibrium (LTE) conditions, the column densities can be written as: 
\begin{equation}
\begin{aligned}
N_{tot}^{thin} = \frac{3h}{8\pi^3S\mu^{2}R_{i}}\frac{Q_{rot}}{g_{u}}\frac{\exp\left(\frac{E_{u}}{kT_{ex}}\right)}{\exp\left(\frac{h\nu}{kT_{ex}}\right)-1}\\
\times \frac{1}{J_{\nu}\left(T_{ex}\right)-J_{\nu}\left(T_{bg}\right)} \int T_{mb}dV,
\end{aligned}
\end{equation}
where {\it S} is the line strength and $\mu$ is the dipole moment. The dipole moments of CCS and {\HCnN} are from \cite{pickett1998} and \cite{deleon1985}, respectively. $E_{u}$ is the upper level energy. For CCS, $R_{i}$ equals 1. For {\HCnN}, the integrated intensity of the two brightest hyperfine components (which are blended together) was used to calculate the column density and $R_{i}$ is calculated from the sum of the hyperfine intensity of the two strongest components ($F=3-2$ and $F=2-1$) divided by the total flux, which corresponds to 0.72. $T_{bg}$ is the background temperature (2.73\,K). We use the excitation temperatures derived from {\NHn} observations. For CCS and {\HCnN}, the degeneracy $g_{u}$ at the energy level {\it u} equals 5. The values of molecule parameters including $\nu$, S$\mu^2$, $E_{u}/k$, and B are from Splatalogue \footnote[1]{https://www.cv.nrao.edu/php/splat/} and are listed in Table~\ref{tab:lines}.

A detailed description of the calculation of the column densities is in \ref{sec:level1}. The calculated parameters including the column densities are listed in Table \ref{tab:fitted_paramters}.

\begin{tablehere}
\caption{Summary of the Parameters for CCS and {\HCnN}.}\label{tab:lines}
\vspace{-1mm}\footnotesize
\begin{center} \doublerulesep 0.3pt \tabcolsep 3.5pt
\begin{tabular}{ c c c c c c  }
\hline
Molecule & Transition  & Frequency (GHz) &
B (GHz)& E$_{u}$/k (K) &  S$\mu^2$ (D$^{2}$) \\
\hline
CCS  & $N\,$=1--0, $J\,$=2--1 & 22.344030 & 6.477 & 1.6 & 16.43 \\
{\HCnN} & $J\,$=2--1 & 18.196279 & 4.549 &1.3 & 27.85 \\
 \hline
\multicolumn{6}{p{.45\textwidth}}{Notes. D equals $1\times 10^{-18}$ esu cm.}
\end{tabular}
\end{center}
\end{tablehere}

\clearpage
\begin{tablehere}
\caption{Fitted Parameters of {\NHn}, CCS, and {\HCnN}.}\label{tab:fitted_paramters}
\vspace{-1mm}\footnotesize
\begin{center} 
\doublerulesep 0.3pt \tabcolsep 10pt
\begin{tabular}{ c c c c c c  c }
\hline
& &  {\NHn}& & & CCS & {\HCnN} \\
Source Name & $T_{rot}$ (K)  & $T_{kin}$ (K)&$T_{ex}$ (K) & $N$($\times10^{14}$\,cm$^{-2}$) & $N$($\times10^{12}$\,cm$^{-2}$) &  $N$($\times10^{12}$\,cm$^{-2}$)\\
\hline
 SDC18.624-0.070 & 12.3 (2.5)& 13.0 (2.6) & 4.8 (0.2) & 3.3 (0.1) & <1.4 & <3.5\\
 SDC18.787-0.286 & 12.8 (1.0) & 13.6 (1.2)& 4.8 (0.1) & 5.5 (0.1) & 2.3 (0.5) & 3.6 (0.4)\\
 SDC18.888-0.476 & 16.2 (0.2)& 17.8 (0.3) & 5.0 (0.1) & 10.8 (0.2) & <0.4 & 3.6 (0.4) \\ 
 SDC21.321-0.139 & 14.6 (2.0)& 15.6 (2.1) & 3.8 (0.2) & 1.9 (0.1) & <0.7 & <0.7\\
 SDC22.724-0.269[1] & 13.2 (1.3)& 14.0 (1.5) & 3.8 (0.1) & 4.5 (0.1) & - & <0.6\\ 
  SDC22.724-0.269[2] & 9.9 (1.3)& 10.2 (1.5) & 3.4 (0.1) & 7.2 (0.2) & 1.7 (0.4) & - \\ 
 SDC23.066+0.049 & 11.6 (4.0)& 12.2 (4.1) & 3.8 (0.1) & 5.8 (0.1) & 0.8 (0.3) & 1.7 (0.5)\\
 SDC23.367-0.288 & 11.3 (2.7)& 11.8 (2.8) & 4.2 (0.3) & 10.8 (0.2) & 2.0 (0.3) & 4.4 (0.5) \\
 SDC24.118-0.175 & 13.4 (1.1)& 14.2 (1.3) & 6.0 (0.1) & 5.1 (0.1) & <1.1 & 5.3 (0.6)\\
 SDC24.433-0.231 & 15.1 (1.0)& 16.4 (1.2) & 5.6 (0.1) & 10.7 (0.2) & <1.1 & 7.0 (0.8) \\
 SDC24.489-0.689 & 13.6 (1.2)& 14.4 (1.4) & 5.0 (0.1) & 6.8 (0.2) & 2.9 (0.4) & 8.0 (0.4)\\
 SDC24.618-0.323& 14.1 (2.0)& 15.0 (2.3) & 4.8 (0.2) & 3.2 (0.1) & <0.8 & 3.0 (0.5) \\
 SDC24.630+0.151 & 13.8 (2.0)& 14.8 (2.3) & 4.8 (0.2) & 4.4 (0.1) & 2.8 (0.6) & 7.4 (0.6)\\
 SDC25.166-0.306 & 13.3 (0.7) & 14.2 (0.8)& 4.4 (0.1) & 2.4 (0.1) & 2.1 (0.4) & <0.5 \\
 SDC25.243-0.447 & 12.9 (1.0) &13.6 (1.1) & 4.6 (0.1) & 3.3 (0.1) & 0.6 (0.2) & 1.6 (0.4)\\
 SDC26.507+0.716$^{(a)}$ & 12.0 (2.5)&12.6 (2.6)  & 4.2 (0.2) & 10.4 (0.2) & 7.1 (2.2) & 5.0 (0.8)\\
 SDC28.275-0.163 & 9.7 (3.0) & 10.0 (3.1) & 4.0 (0.3) & 9.1 (0.2) & <1.8 & <0.5\\
 SDC28.333+0.063 & 11.3 (1.3) & 11.8 (1.4) & 5.4 (0.2) & 5.3 (0.1) & 3.1 (0.4) & 6.6 (0.5)\\
SDC31.039+0.241 [1] & 10.2 (1.3) &  10.6 (1.4)& 5.4 (0.2) & 5.3 (0.1) & <0.7 & 2.0 (0.5)\\
 SDC31.039+0.241 [2] & 10.6 (1.3)&11.0 (1.4)  & 5.0 (0.2) & 6.2 (0.2) & - & 2.0 (0.7)\\ 
 SDC34.370+0.203 & 12.1 (4.0)& 12.6 (4.3) & 5.4 (0.4) & 11.7 (0.2) & <2.7 & <2.2 \\
 SDC35.429+0.138 & 12.4 (1.5)& 13.0 (1.7) & 5.2 (0.2) & 2.9 (0.1) & 1.6 (0.2) & 2.9 (0.2)\\
 SDC35.527-0.269 & 11.6 (1.2) & 12.2 (1.3) & 5.6 (0.1) & 3.1 (0.1) & 2.1 (0.3) & 2.5 (0.3) \\
 SDC35.745+0.147 & 12.9 (0.8)&13.6 (1.0)  & 6.6 (0.1) & 5.7 (0.2)  & 1.9 (0.4) & 4.9 (0.5)\\
 SDC38.850-0.427 & 15.9 (2.0)& 17.4 (2.2) & 6.2 (0.2) & 1.8 (0.1) & <0.5 & 2.4 (0.4)\\
 SDC40.283-0.216 & 13.8 (2.0)& 14.8 (2.3) & 4.4 (0.2) & 7.5 (0.2) & <1.2 & 3.6 (0.6)\\
 SDC47.061+0.257$^{(a)}$ & 15.9 (3.0)& 17.4 (3.2) & 3.6 (0.2) & 4.4 (0.1) & 2.6 (0.7) & <1.1\\
 SDC52.723+0.045 & 14.8 (3.0)& 16.0 (3.2) & 4.2 (0.4) & 0.5 (0.1) & <0.6 & 1.6 (0.6)\\
\hline
\multicolumn{7}{p{.9\textwidth}}{Notes. The values in the parentheses are the uncertainties. $^{(a)}$ The derived parameters for these two sources (SDC26.507+0.716 and SDC47.061+0.257) are from one velocity component. These two sources have two velocity components for {\NHn}, which are heavily overlapping. For CCS and {\HCnN}, the low signal-to-noise ratios limit the reliability of fitting two velocity components related to {\NHn} velocities. We exclude these two sources from further analysis.}
 \end{tabular}
\end{center}
\end{tablehere}
\clearpage

\section{Discussion}\label{sec:discussion}

\subsection{Possible evolutionary scenario}\label{subsec:chemical_evolution}
The correlation and anti-correlation of CCS, {\HCnN}, and {\NHn} have been attributed to the connection among the formation of hydrogen-, nitrogen- and sulphur-terminated CCMs, which has been established in low-mass star-forming regions \cite{hirahara1992}. The abundance of CCS is found to peak in chemically young gas and then decrease due to its being processed into molecules such as CO \cite{devine2011}.
The lifetime of CCS is suggested to be only 10$^{4}$ years after the onset of star formation \cite{gregorio2006}. However, the collisions of subclouds can trigger waves or shocks, which, even at low speeds, could reset the chemical clock and lead to the resurrection of early-time species like CCS\cite{dickens2001}. Hydrogenated molecules, such as {\NHn} and {\HCnN}, may be enhanced in star-forming regions as ice evaporates off dust grains (e.g.,\cite{vandishoeck1998,taniguchi2018-hmpo}). The modelled abundance of {\HCnN} first increases strongly at early times and decreases rapidly after 10$^5$ yr due to C+HC$_n$N reactions\cite{vasyunina2012}. It is clear that the column density ratios between N-bearing species and carbon-chain species are sensitive to the chemical environment and can potentially be used as chemical evolutionary indicators in star-forming regions\cite{suzuki1992,benson1998,aikawa2001,tatematsu2014}.

To investigate the chemical evolutionary stages, we adopt the three-phase modelling code {\sc NAUTILUS} \cite{ruaud2016} 
to compute the evolution of chemical abundances with the physical parameters of the sources and compare the results with the observation. {\sc NAUTILUS} includes the gas-phase reactions, interactions between species in the gas-phase and grain surfaces, and chemical reactions at the surface of grains\cite{ruaud2016}. The gas-phase network is based on public network data KInetic Database for Astrochemistry (KIDA)\footnote{http://kida.astrophy.u-bordeaux.fr/} \cite{wakelam2015}. The model includes the direct photodissociation by photons and cosmic-ray induced secondary UV photons\cite{prasad1983}, the competition between reaction, diffusion and evaporation, as suggested by \cite{chang2007}, and several non-thermal desorption mechanisms such as the cosmic-ray desorption mechanism\cite{hasegawa1992}, the chemical desorption mechanism\cite{garrod2007}, and photo-desorption\cite{wakelam2015}.

 We introduced both low- and high-metal abundances models to simulate the chemical evolution for conditions of metals being  heavily depleted and not, respectively. The initial elemental abundances are listed in Table \ref{tab:initial_abundances}. In the low-metal elemental abundances setting, S, Si, Mg, Fe, Na, Cl, and P are depleted by two orders of magnitude with respect to the high-metal elemental one, as those metals are suggested to be more depleted in dense clouds or even totally depleted on grains \cite{graedel1982,flower2003, wakelam2008}. Currently, our knowledge of the chemical processes in the IRDCs is insufficient to determine which set of abundances is more applicable. We thus present both low- and high-metal models for IRDCs\cite{vasyunina2012,beaklini2020}. We use an updated Sulphur chemistry \cite{vidal2017}, which contains 1139 species and 13227 reactions, for both low- and high-metal abundances models. Since single-dish observations limit the resolution of the structure locating at several kpc to no smaller than 0.5\,pc, uniform temperature and density are suitable for the chemical simulations. We list the parameters in the chemical models in Table \ref{tab:chemical_model}. The parameters are all typical values for massive dark clouds \cite{vasyunina2012}.

\noindent
\begin{table}
\mbox{}
\begin{minipage}[t]
{.47\textwidth}
\vspace{-1mm}\footnotesize
\begin{center} \doublerulesep 0.3pt \tabcolsep 3 pt
\caption{Initial Elemental Abundances with Respect to nH$^{(a)}$}\label{tab:initial_abundances}
\begin{tabular}{ c  c  c  }
\hline
Element & Low-metal Abundance$^{(b)}$& High-metal Abundance$^{(c)}$ \\
\hline
He & 9.00(-2) & 9.00 (-2)\\
N & 7.60 (-5)& 7.60 (-5)\\
O & 2.56 (-4)& 2.56 (-4)\\
C$^{+}$ & 1.2 (-4)& 1.20 (-4) \\
S$^{+}$ & 8.00 (-8)& 1.50 (-5)\\
Si$^{+}$ & 8.00 (-9)& 1.70 (-6) \\
Fe$^{+}$ & 3.00 (-9) & 2.00 (-7)\\
Na$^{+}$ & 2.00 (-9) & 2.00 (-7)\\
Mg$^{+}$ & 7.00 (-9) & 2.40 (-6)\\
Cl$^{+}$ & 1.00 (-9) & 1.8 (-7)\\
P$^{+}$ & 2.00 (-10) & 1.17 (-7)\\
 \hline
\multicolumn{3}{p{.8\textwidth}}{Notes. $^{(a)}$ nH refers to the number density of hydrogen nuclei. {\it a(b)} denotes a $\times$ 10$^{b}$. $^{(b)}$ This set is from \cite{semenov2010} and similar to EA 1 from \cite{wakelam2008}. $^{(c)}$ This set is from \cite{wakelam2008}. The differences in elemental abundances of C, N, and O are within a factor of $\sim$2 in all models \cite{flower2003,wakelam2008}}.
\end{tabular}
\end{center}
\end{minipage}
\begin{minipage}[t]
	{.45\textwidth}
\vspace{6mm}\footnotesize
\begin{center} \doublerulesep 0.3pt \tabcolsep 3 pt
\caption{Parameters of the Chemical Models}\label{tab:chemical_model}
\begin{tabular}{ c c }
\hline
 Parameter & Value\\
\hline
nH$^{(a)}$ & 8$\times$10$^{4}$\,(cm$^{-3}$) to 1$\times$10$^{7}$\,(cm$^{-3}$)\\
Gas/dust temperature & 15\,K\\
Cosmic-ray ionization rate & 1.3 $\times$ 10$^{-17}$ s$^{-1}$\\
Grain radius & 0.1 $\mu$m\\
Grain density & 3.0 g cm$^3$\\
Diffusion/desorption energy ratio & 0.4\\
Gas-to-dust mass ratio & 100\\
Extinction & 16\\
Surface site density & 1.5 $\times$ 10$^{15}$ cm$^{-2}$\\
 \hline
 \multicolumn{2}{p{.8\textwidth}}{Notes. $^{(a)}$ nH refers to the number density of hydrogen nuclei.}
\end{tabular}
\end{center}
\end{minipage}
\mbox{}
\end{table}

To evaluate the goodness of fit, we calculated the mean logarithmic difference using the logarithmic parameter defined by $\sum | \log(X_{mod}) - \log(X_{obs}) |/N_{obs}$ \cite{loison2014}, where X$_{mod}$ and X$_{obs}$ are the modelling and observational abundance, respectively. The best fit is then found with the minimum of this logarithmic parameter. We adopted an order of magnitude, i.e. (X$_{obs}/10\leq$ X$_{model}\leq$ 10X$_{obs}$), as the range to search for the best fit. This criterion has been widely adopted in astrochemical simulations\cite{wakelam2008}, as well as in the comparisons with observational data (e.g. \cite{vasyunina2012,suzuki2014}).

\begin{figure}[H]
\centering
\includegraphics[width=0.50\textwidth,height=0.35\textwidth, clip]{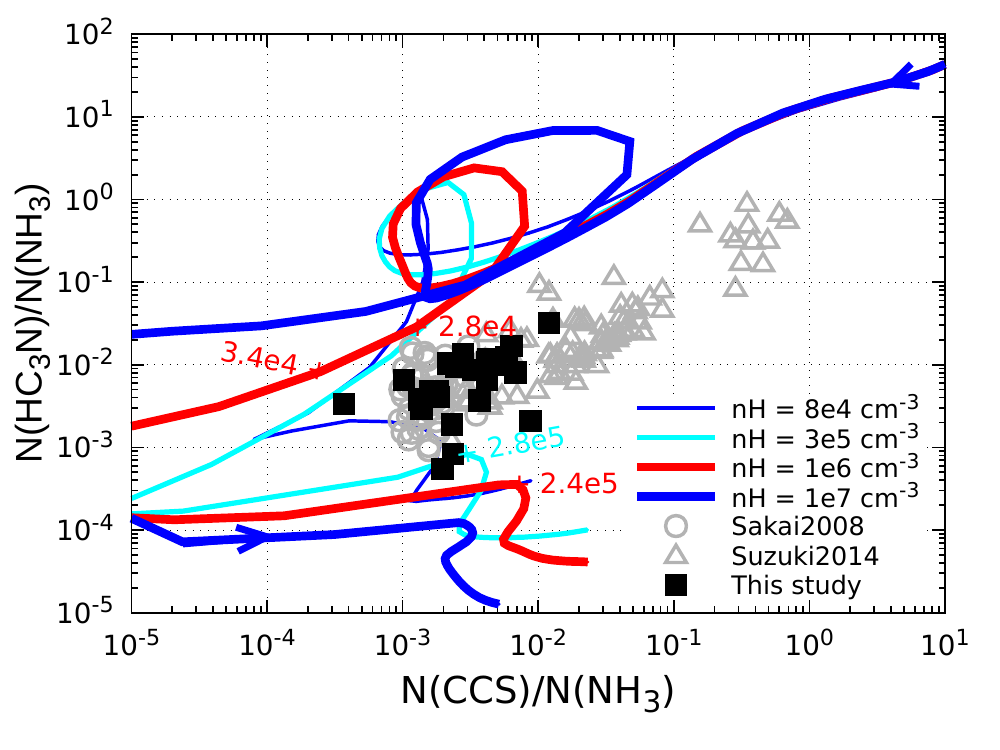}
\includegraphics[width=0.50\textwidth,height=0.35\textwidth, clip]{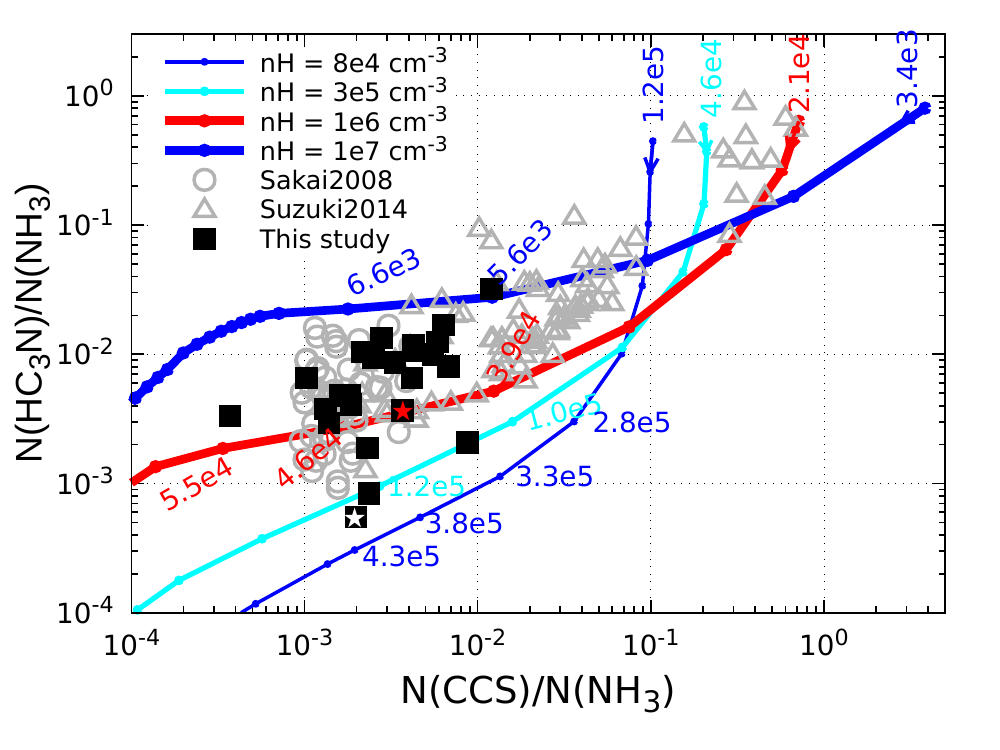}
\caption{{\it N}(CCS)/{\it N}({\NHn}) versus those of {\it N}({\HCnN})/{\it N}({\NHn}) overlaid with the chemical evolutionary tracks predicted by low- (upper panel) and high- (lower panel) metal abundance models for different densities ranging from 8$\times$10$^{4}$ to 1$\times$10$^{7}$\,{cm$^{-3}$}. In both panels, the arrows show the direction of the evolution. The values along each line represent the chemical times at the corresponding points. The filled black squares are the data from this TMRT survey. The open gray circles are the data observed towards a sample of 16 IRDCs \cite{sakai2008}. The open gray triangles represent the data towards low-mass star-forming regions \cite{suzuki1992,hirota2009,suzuki2014}. Upper: The evolutionary tracks for these densities compass this TMRT observed sample and do not match the majority of the sample. The best fit data point provides an upper limit of 2.8$\times$10$^{5}$\,yr. Lower: The evolutionary tracks can cover this TMRT observed sample, with the lowest and highest densities being the lower and upper boundary, respectively. The red star represents source SDC21.321, which fitted by the evolutionary track of density 1$\times$10$^{6}$\,{cm$^{-3}$} at 4.1$\times$10$^{4}$\,yr. The white star represents source SDC28.275, which is the data point closest to the evolutionary track at density 8$\times$10$^{4}$\,{cm$^{-3}$}.
}
\label{fig:stages}
\end{figure}

The evolutionary tracks for different cloud densities, of column density ratios of {\HCnN} to {\NHn} (denoted as {\it N}({\HCnN})/{\it N}({\NHn}) hereafter) versus those of CCS to {\NHn} (denoted as {\it N}(CCS)/{\it N}({\NHn}) hereafter), from both low- and high-metal abundances modellings, are presented in the upper and lower panels of Figure~\ref{fig:stages}, respectively. For the low abundance models the column density ratios are not unique functions of the time in the models. The observed range of {\it N}(CCS)/{\it N}({\NHn}) occurs for two age ranges in the models but none of them have {\it N}({\HCnN})/{\it N}({\NHn}) consistent with the majority of the observations. In contrast to the low abundance models, the high abundance models do show a one-to-one correspondence between column density ratios and the evolutionary timescale of a source. The high-metal models can bound the range of the observed column density ratios for densities from 8$\times10^{4}$\,cm$^{-3}$ to 1$\times10^{7}$\,cm$^{-3}$, although this upper limit is probably much higher for the density sampled by this survey. The implied evolutionary timescale is a function of the density, ranging from $\sim4.3\times10^{5}$\,yr for a density of 8$\times10^{4}$\,cm$^{-3}$ to $\sim6.6\times10^{3}$\,yr for a density of 1$\times10^{7}$\,cm$^{-3}$. The dispersion in the sample in Figure~\ref{fig:stages} could be raised from the various physical properties of the sources, but may also be caused by the different evolutionary times of the sources. If the latter is the case, our model shows that the timescale for this IRDC sample is $\lesssim$ 1$\times$10$^{5}$\,yr. The higher density models tend to fit lower ages, indicating that the chemistry of denser regions evolve faster than less dense regions.

\begin{figure}[H]
\centering
\includegraphics[width=0.50\textwidth,height=0.35\textwidth, clip]{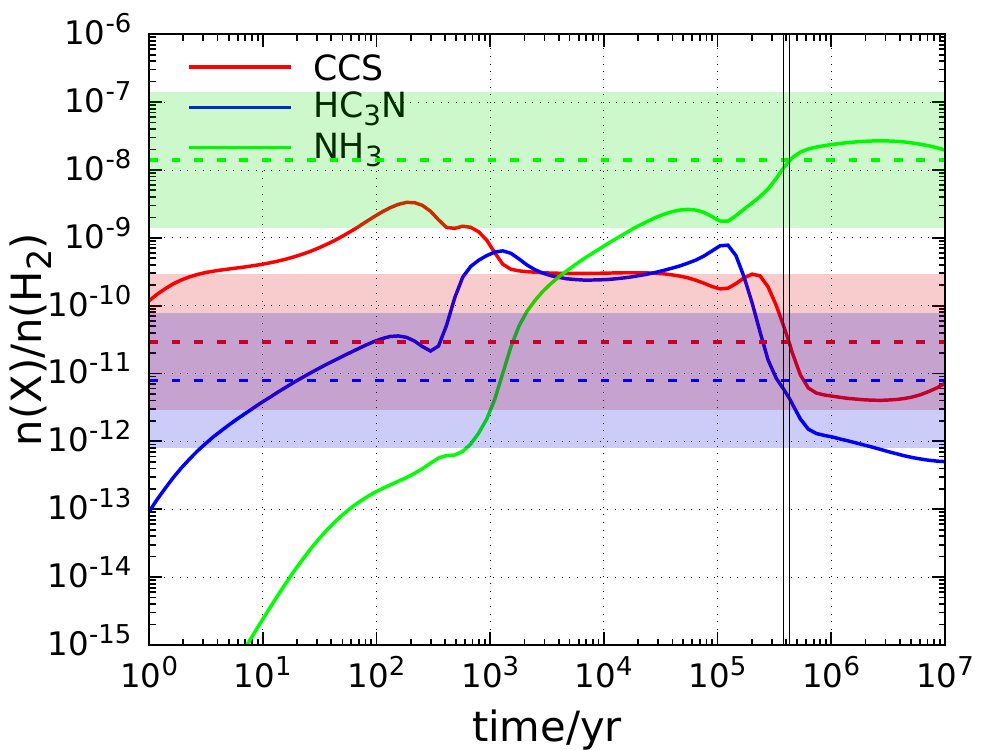}
\includegraphics[width=0.50\textwidth,height=0.35\textwidth, clip]{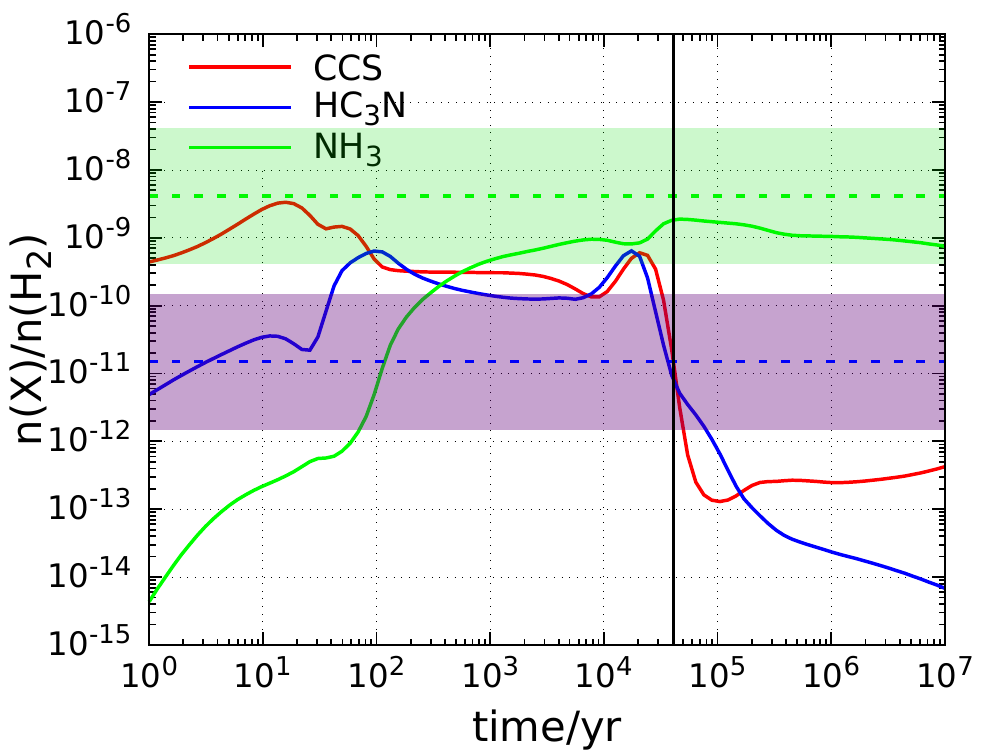}
\caption{The modelling results of these molecular fractional abundances to H$_{2}$ as a function of evolutionary timescales from the high-metal abundances simulations. Upper: the modelling result at the density of 8$\times10^{4}$\,{cm$^{-3}$} for source SDC28.275, labelled as the white star in Figure~\ref{fig:stages}. The vertical black lines represent 3.8$\times10^{5}$\,yr and 4.3$\times10^{5}$\,yr. The best fit time (3.9$\times10^{5}$\,yr) is in between of these two lines. Lower: the modelling result at the density of 1$\times10^{6}$\,{cm$^{-3}$} for source SDC21.321, labelled as the red star in Figure~\ref{fig:stages}. The dashed lines represent the abundance of the corresponding molecules of the same colours. The stripes represent the two magnitudes range to search for the best fit. The vertical black line represents the best fit time.}
\label{fig:bestfit}
\end{figure}

The more commonly used abundance vs.\ time plots are presented in Figure~\ref{fig:bestfit}, the upper panel of which is for source SDC28.275, lower panel for source SDC21.321. With the two orders of magnitude range criteria, the models have the time range which match with the observed fractional abundances of the three molecules for both sources, as indicated by the coloured stripes in Figure~\ref{fig:bestfit}. For source SDC28.275, labelled as the white star in the lower panel in Figure~\ref{fig:stages}, falls close to, but not on the evolutionary track of density 8$\times10^{4}$\,cm$^{-3}$. This can be explained by the fractional abundance plot modelled for the same density in the upper panel of Figure~\ref{fig:bestfit}, where {\it N}({\HCnN})/{\it N}({\NHn}) and {\it N}(CCS)/{\it N}({\NHn}) at each vertical black line can fit for only one column density ratio of the source at one time, but no suitable timescale can fit for both column density ratios. For source SDC21.321, indicated by the red star at the evolutionary track of density 1$\times10^{6}$\,cm$^{-3}$ in the lower panel in Figure~\ref{fig:stages}, all the fractional abundances from the model are smaller than the observational data, but the column density ratios from the model can match with the observed column density ratios. Though this density (1$\times10^{6}$\,cm$^{-3}$) is higher than the average density of the cloud, it could still be plausible as the dense clumps, where the massive stars and their associated clusters form, could have a density up to 1$\times10^{8}$ cm$^{-3}$ \cite{peretto2013}, considering all molecules in this study are dense gas tracers. We note that illustrations such as Figure~\ref{fig:bestfit} might be limited and the column density ratios of molecules such as Figure~\ref{fig:stages} can provide more detailed comparisons of estimating the chemical evolutionary timescales, which for this sample are $\lesssim$10$^{5}$\,yr.

\subsection{Comparisons with Other Studies}\label{subsec:comparison_previous}

The low detection rate of CCS has been believed to be an indication that these regions are more chemically evolved \cite{sakai2008,taniguchi2019-hmpo}. The first attempt to observe CCS towards IRDCs was carried out using single-pointing observations of the $J_{N}$\,=4$_{3}$-3$_{2}$ transition \cite{sakai2008}, which is at higher energy than the $J_{N}$\,=2$_{1}$-1$_{0}$ transition observed here. However, a combined GBT and VLA survey\cite{dirienzo2015} achieved higher detection rates of CCS in $J_{N}$\,=2$_{1}$-1$_{0}$, suggesting that the detection rate of a molecular depends on the excitation energy needed for the transition observed. The upper limit on the CCS column density derived from the $J_{N}$\,=4$_{3}$--3$_{2}$ survey was a few $\times$10$^{12}$\,{cm$^{-3}$}\cite{sakai2008}, which is similar to the values measured in this TMRT study. This points to the need to consider excitation effects when considering detection rates and their implications for the evolutionary stages of regions.

The temperatures and line widths from this TMRT {\NHn} observation agree with other large sample IRDC ammonia surveys made by Effelsberg telescope, GBT, VLA, and Parkes telescope \cite{pillai2006,hill2010,ragan2011,chira2013} and other massive star-forming regions, such as Orion\cite{li2013}. The mean full width half maximum (FWHM) of {\NHn} in this sample is larger than that of low-mass star-forming regions \cite{jijina1999,friesen2009} but smaller than that of High Mass Protostellar Objects (HMPOs)\cite{pillai2006,chira2013}. HMPO has a central protostar with mass $>$8 {\Msun} which is still accreting (e.g. \cite{sridharan2002,beuther2007}). It is considered that HMPOs are more evolved than IRDCs and are the next evolutionary step after IRDCs in massive star formation\cite{beuther2007}. So the smaller FWHM of {\NHn} can be considered as an additional evidence that IRDCs are at the earliest stages of forming high mass stars though not all of them become massive stars\cite{chira2013}.

As shown in Figure~\ref{fig:stages}, {\it N}(CCS)/{\it N}({\NHn}) and {\it N}({\HCnN})/{\it N}({\NHn}) are consistent among IRDCs. It also shows that although there is a large scatter in {\it N}(CCS)/{\it N}({\NHn}), it is this ratio that most separates the IRDCs from the CCS-rich low-mass sources. For the IRDCs, {\it N}(CCS)/{\it N}({\NHn}) $\lesssim10^{-2}$ whereas for the CCS-rich low-mass sources {\it N}(CCS)/{\it N}({\NHn}) $\gtrsim0.5\times10^{-2}$ and almost all the low-mass sources have $\gtrsim10^{-2}$. Compared with the most CCS-rich low-mass sources (those with {\it N}(CCS)/{\it N}({\NHn}) > 0.1), the ratio {\it N}({\HCnN})/{\it N}({\NHn}) observed in the IRDCs is also $\sim 10^{2}$ smaller. However, the majority of the low-mass sources have $2\times10^{-3}<${\it N}(\HCnN)/{\it N}(\NHn)$<3\times10^{-2}$ which are similar to the majority of the IRDCs. The IRDC closest to the CCS-rich low-mass sources has the lowest mass (258\,{\Msun}), though no correlation has been found in either ratio with the mass of IRDCs. The much lower ratio of $N$(CCS)/$N$({\NHn}) in the IRDCs, which contributed mostly by the greater column densities of {\NHn}, could point to major differences in their evolution compared with low-mass star-forming regions.

\section{Conclusion}

We describe here one of the first K Band spectroscopic observations of the TMRT. We successfully detected {\NHn}(1,1) and (2,2) towards all 26 IRDCs in our sample. Two carbon-chain molecules CCS and {\HCnN} were simultaneously observed with the K Band receiver. Our main conclusions are summarized as follows.

1. The detection rates of {\HCnN} (73\%) and {\NHn} (100\%) are similar to the observations towards other samples of IRDCs, while the total number of the existing CCS-detected IRDCs has been increased drastically from 8 to 23 with this survey.

2. The derived IRDC gas temperatures from {\NHn} inversion lines, using the newly developed HGFR method, range from 10 to 18\,K, consistent with their tracing the early phases of massive star formation and the stellar cluster formation.

3. The similarities in line widths among CCS, {\HCnN}, and {\NHn} suggest that these species trace similar regions.

4. In the IRDCs, $N$(CCS)/$N$({\NHn}) are between 3.7$\times 10^{-4}$ and 1.5$\times 10^{-2}$, whereas $N$({\HCnN})/$N$({\NHn}) are between 5.5$\times 10^{-4}$ and  3.2$\times 10^{-2}$. In the low-mass star-forming regions, the ranges for $N$(CCS)/$N$({\NHn}) and $N$({\HCnN})/$N$({\NHn}) are from 2.1$\times 10^{-3}$ to 6.9$\times 10^{-1}$ and 1.3$\times 10^{-3}$ to 9.0$\times 10^{-1}$, respectively. The larger  $N$(CCS)/$N$({\NHn}) and $N$({\HCnN})/$N$({\NHn}) in the low-mass star-forming regions than those ratios in the IRDCs are mainly due to the smaller amount of {\NHn} in low-mass star-forming regions. The ratios between carbon chains and Nitrogen-bound carbonless molecules are shown to be a clear differentiating factor between low- and high-mass star forming regions. 

5. We model our observations using a three-phase reaction network, {\sc NAUTILUS}, with both low- and high-metal initial elemental abundances. The high-metal abundance models better reproduce the column density ratios of CCMs to {\NHn}. 
Our best fit models for these IRDCs are all consistent with their being younger than 10$^{5}$\,yr, similar to the chemical ages found in deuteration studies towards massive star-forming regions\cite{gerner2015}.

\vspace*{2mm} \Acknowledgements{\bahao This work was supported by National Natural Science Foundation of China (NSFC) (Grant Nos. 11988101, 11725313, 11911530226, and 11403041), and the Chinese Academy of Sciences (CAS) International Partnership Program (Grant No. 114A11KYSB20160008). The authors would like to thank the referees for the many constructive comments.
XIE J. J. would like to thank the support and help from the TMRT operating team and the co-observer Y. T. Yan during her observations at Shanghai Tian Ma Radio Telescope. XIE J. J. would also like to thank Xintong Lu and Yongxiong Wang for correcting the English in this paper.
G. A. F acknowledges financial support from the State Agency for Research of the Spanish MCIU through the AYA2017-84390-C2-1-R grant (co-funded by FEDER) and through the ``Center of Excellence Severo Ochoa'' award for the Instituto de Astrof\'isica de Andalucia
(SEV-2017-0709). G.A.F also acknowledges support from the Collaborative Research Centre 956, funded by the Deutsche Forschungsgemeinschaft (DFG) project ID 184018867.}

\clearpage
\begin{appendix}




\renewcommand{\thesection}{Appendix}

\section{}






\subsection{\label{sec:level1} Column Density Determination}

\renewcommand{\theequation}{a\arabic{equation}}
\setcounter{equation}{0}

The excitation temperature, T$_{ex}$ in Eq. \ref{eq:n11}, is related to T$_{mb}$ through:
\begin{equation}\label{eq:a1}
\begin{aligned}
T_{mb}\left(\nu\right)=\eta_{f}\left[J_{\nu}\left(T_{ex}\right)-J_{\nu}\left(T_{bg}\right)\right]\left[1-e^{-\tau\left(\nu\right)}\right],
\end{aligned}
\end{equation} where $\eta_{f}$ is the main beam filling factor. Assuming that the emission from the molecular lines is extended and uniformly distributed within the beam, the beam filling factor~{$\eta_{f}$} is thus $\sim$1. $J_{\nu}T$ is the {\it Rayleigh-Jeans equivalent temperature}, which is the equivalent temperature of a black body at temperature {\it T}: 
\begin{equation}\label{eq:a2}
\begin{aligned}
J_{\nu}\left(T\right)\equiv \frac{\frac{h\nu}{k}}{\exp\frac{h\nu}{kT}-1},
\end{aligned}
\end{equation}.

As CCS has two unpaired electrons, for CCS and {\HCnN} column density calculation, we calculate the partition function $Q_{rot}$ for a state with energy above the ground state $E_{i}$ and degeneracy $g_{i}$ using:
\begin{equation}\label{eq:a6}
	Q_{rot} = \sum_{i} g_{i}\exp\left(-\frac{E_{i}}{kT_{k}}\right), 
\end{equation}
adopting the values of $E_{i}$ and $g_{i}$ from the tabulated data for CCS in the JPL molecular spectroscopy catalog \cite{pickett1998} and assuming that 300 states are thermally populated. The value of the partition function varies between 18.33 at 4\,K and 60.23 at 10\,K. For {\HCnN}, the approximation of the partition function $Q_{rot}$ is:
\begin{equation}\label{eq:a7}
	Q_{rot} \simeq \frac{kT}{hB}\exp\left(\frac{hB}{3kT}\right),
\end{equation}
where B is the rigid rotor rotation constant.

\end{appendix}













\end{document}